\documentclass[prc,preprint,showpacs,superscriptaddress,floatfix,amsmath,amssymb,nofootinbib]{revtex4}
\usepackage{graphicx}
\usepackage{color}
\usepackage{CJK}
\usepackage{ulem}

\newcommand{\beq}{\begin{equation}}
\newcommand{\eeq}{\end{equation}}
\newcommand{\beqn}{\begin{eqnarray}}
\newcommand{\eeqn}{\end{eqnarray}}

\newcommand{\Lb}{\left\{}
\newcommand{\Rb}{\right\}}
\newcommand{\lb}{\left(}
\newcommand{\rb}{\right)}

\begin{document}
\begin{CJK*}{GBK}{song}
\title{Chirality in odd-$A$ Rh isotopes within triaxial particle rotor model}
\author{B. Qi}
\affiliation{School of Space Science and Physics, Shandong
University at Weihai, Weihai 264209, China}
\author{S.Q. Zhang}\thanks{e-mail: sqzhang@pku.edu.cn}
\affiliation{State Key Laboratory of Nuclear Physics and Technology,
School of Physics, Peking University, 100871 Beijing, China }
\author{S.Y. Wang}
\affiliation{School of Space Science and Physics, Shandong
University at Weihai, Weihai 264209, China}
\author{J. Meng}
\affiliation{School of Physics and Nuclear Energy Engineering,
Beihang University, Beijing 100191, China}
\affiliation{State Key
Laboratory of Nuclear Physics and Technology, School of Physics,
Peking University, 100871 Beijing, China }
 \affiliation{ Department
of Physics, University of Stellenbosch, Stellenbosch, South Africa}
\author{T. Koike}
\affiliation{Department of Physics, Tohoku University, Sendai
980-8578, Japan}
\date{\today}

\begin{abstract}
Adopting the fully quantal triaxial particle rotor model, the
candidate chiral doublet bands in odd-$A$ nuclei $^{103}$Rh and
$^{105}$Rh with $\pi g_{9/2}^{-1}\otimes\nu h^{2}_{11/2}$
configuration are studied. For the doublet bands in both nuclei,
agreement is excellent for the observed energies over entire spin
range and $B(M1)/B(E2)$ at higher spin range. The evolution of the
chiral geometry with angular momentum is discussed in detail by the
angular momentum components and their probability distributions.
Chirality is found to change from  chiral vibration to nearly static
chirality at spin $I=37/2$ and back to another type of chiral
vibration at higher spin. The influence of the triaxial deformation
$\gamma$ is also studied.

\end{abstract}


\maketitle

\section{Introduction}
Since the theoretical prediction of spontaneous chiral symmetry
breaking in nuclear structure in 1997~\cite{FM97}, much effort has
been devoted to further explore this interesting phenomenon. So far,
more than 20 candidate chiral nuclei have been reported
experimentally in the $A\sim100$, 130 and 190 mass regions
~\cite{Starosta01,Koike01,Bark01,Mergel02, Koike03, Zhu03, Joshi04,
Alcantara04, Timar04, Vaman04, Timar06, Tonev06, WangSY06a,
Lawrie08,Joshi07,Grodner06}. An overview of these studies and open
problems in understanding the nuclear chirality is introduced in
Ref.~\cite{Meng10}.

Of particular interest is Rh isotope chain, which has attracted
significant attentions. Not only in odd-odd
$^{102}$Rh~\cite{Timar06}, $^{104}$Rh~\cite{Vaman04},
$^{106}$Rh~\cite{Joshi04}, but also in odd-$A$
$^{103}$Rh~\cite{Timar06} and $^{105}$Rh~\cite{Timar04}, the
existence of chiral doublet bands has been claimed, while lifetime
measurements for candidate chiral bands in $^{103,104}$Rh have been
performed using the recoil distance Doppler-shift
method~\cite{Suzuki08}. The corresponding quasiparticle
configurations are suggested as $\pi g_{9/2}^{-1}\otimes \nu
h_{11/2} $ for odd-odd isotopes and $\pi g_{9/2}^{-1}\otimes \nu
h^2_{11/2} $ for odd-$A$ isotopes. In particular, the candidate
chiral bands in neighboring four nuclei $^{102,103,104,105}$Rh have
been regarded as a special quartet of nuclear
chirality~\cite{Timar06}. The study on this quartet permits us to
investigate the detail influence on the nuclear chirality of valence
nucleon configurations and inert core and to testify the validity of
the theoretical approaches.

On the theoretical side, various approaches have been applied to
study the nuclear chirality in Rh
isotopes~\cite{Frauendorf01,Meng08MPA,Meng10}. Three-dimensional
tilted axis cranking (TAC) approximation~\cite{Dimitrov00PRL} has
been performed for the doublet bands with $\pi g_{9/2}^{-1}\otimes
\nu h^2_{11/2} $ configuration in $^{105}$Rh, and the aplanar
solutions are obtained at relatively high rotation
frequencies~\cite{Timar04}. Using the quantal particle-rotor model
(PRM), in which two quasiparticles are coupled to a triaxial
core~\cite{Zhang07}, the observed energy spectra, energy staggering
parameter, and electromagnetic transition ratios of the candidate
chiral doublet bands in $^{106}$Rh were well reproduced, which
supports their chiral interpretation~\cite{WangSY08}. Furthermore,
the adiabatic and configuration-fixed constrained triaxial
relativistic mean field approaches have been applied to study the
quasiparticle configurations and the corresponding triaxial
deformation in $^{106}$Rh~\cite{MengJ06} as well as other Rh
isotopes~\cite{PengJ08}, and an interesting phenomenon, multiple
chiral bands (M$\chi$D), is suggested in $^{104,106,108,110}$Rh. The
prediction of M$\chi$D has been further examined by including
time-odd fields in relativistic mean field~\cite{YaoJM09}.

However, for odd-$A$ nuclei $^{103}$Rh~\cite{Timar06} and
$^{105}$Rh~\cite{Timar04},  the lack of nuclear model capable of
addressing their doublet band separations has prevented the previous
observation~\cite{Timar06} from being compared quantitatively with
expectations for the chiral scenario. Such quantitative comparison
for the separations between doublet bands can be carried out by
either triaxial PRM with $n$-particle-$n$-hole
configurations~\cite{Qi0901} or random phase approximation (RPA)
calculations based on TAC mean field~\cite{Mukhopadhyay07}. The
triaxial $n$-particle-$n$-hole PRM was recently developed and has
been applied to study the nuclear chirality in odd-$A$ nucleus
$^{135}$Nd~\cite{Qi0901}. For the chiral doublet bands in
$^{135}$Nd, the observed energies and the electromagnetic
transitions are reproduced excellently, and chirality is found to
change from chiral vibration to nearly static chirality at spin
$I=39/2\,\hbar$ and back to another type of chiral vibration at
higher spin. It is therefore interesting to further study the
doublet bands in odd-$A$ Rh isotopes in the framework of the fully
quantal 2-particle-1-hole coupled to triaxial rotor model, which
could compare experimental data quantitatively with expectations for
the chiral scenario and would shed a new light on the study of
nuclear chirality.

In this paper, the candidate chiral doublet bands in odd-$A$ nuclei
$^{103,105}$Rh will be studied via a triaxial $n$-particle-$n$-hole
PRM. The energy spectra and the electromagnetic transition ratios of
the doublet bands in $^{103, 105}$Rh will be calculated and compared
with the available data. Their chiral geometry will be discussed.
Furthermore, we will investigate the influence of the triaxial
deformation $\gamma$ on the properties of the doublet bands.

\section{Formalism}

The total Hamiltonian is expressed as,
 \beq\label{eq:multiPRM}
  \hat H= \hat H_\textrm{coll}+ \hat H_\textrm{intr},
 \eeq
with the collective rotor Hamiltonian $H_\textrm{coll}$,
 \beq
 \hat H_\textrm{coll}=\sum_{k=1}^3
 \frac{\hat R_k^2}{2{\cal{J}}_k}=\sum_{k=1}^3
 \frac{(\hat I_k-\hat J_k)^2}{2{\cal{J}}_k},
 \eeq
where the indices $k=1,2,3$ refer to the three principal axes of the
body-fixed frame, $\hat{R}_k, \hat{I}_{k}, \hat{J}_{k}$ denote the
angular momentum operators for the core, the total nucleus and the
valence nucleons, respectively. The moments of inertia for
irrotational flow are adopted, i.e., ${\cal J}_k = {\cal
J}_0\sin^2(\gamma - {2\pi k}/{3})$. The intrinsic Hamiltonian for
valence nucleons is
 \beq \label{eq:sp}
 \hat H_\textrm{intr}
  = \sum_{\nu}\varepsilon_{p,\nu}
  a_{p,\nu}^{+}a_{p,\nu}
  + \sum_{\nu'}\varepsilon_{n,\nu'}
  a_{n,\nu'}^{+}a_{n,\nu'}.
 \eeq
The single particle energy for proton $\varepsilon_{p,\nu}$ and for
neutron $\varepsilon_{n,\nu'}$ are obtained by the diagonalization
of the triaxial deformed single-$j$ shell Hamiltonian~\cite{FM97},
 \beq\label{eq:hsp}
  h_\textrm{sp}=\pm \frac{1}{2}C
  \Lb\cos\gamma(\hat j_3^2-\frac{j(j+1)}{3})
  + \frac{\sin\gamma}{2\sqrt{3}}(\hat j_+^2+\hat j_-^2)\Rb,
 \eeq
where the plus or minus sign refers to particle or hole, and the
coefficient $C$ is proportional to the quadrupole deformation
$\beta$ as in Ref.~\cite{WangSY08}.

The single particle states are expressed as
 \beq \label{eq:spwf}
 {a}^{+}_{\nu}| 0 \rangle
 =\sum_{\alpha \Omega}c_{\alpha \Omega}^{(\nu)}
 |\alpha,\Omega \rangle,
    ~~~~
 {a}^{+}_{\overline{\nu}}| 0 \rangle
 = \sum_{\alpha \Omega}(-1)^{j-\Omega}c_{\alpha \Omega }^{(\nu)}
 |\alpha,-\Omega \rangle,
 \eeq
where $\Omega$ is the projection of the single-particle angular
momentum ${\hat j}$ along the 3-axis and is restricted to the values
$\cdots, -7/2, -3/2, 1/2, \cdots$, due to the time-reversal
degeneracy~\cite{Larsson78}, and $\alpha$ denotes the other quantum
numbers. For a system with $z$ valence protons and $n$ valence
neutrons, the intrinsic wave function is given as
 \beq\label{eq:configurations}
 |\varphi\rangle = \lb\prod_{i=1}^{z_{1}}a^\dag_{p, \nu_i}\rb
 \lb\prod_{i=1}^{z_{2}}a^\dag_{p,\overline{\mu_i}}\rb
 \lb\prod_{i=1}^{n_{1}}a^\dag_{n,\nu'_i}\rb
 \lb\prod_{i=1}^{n_2}a^\dag_{n,\overline{\mu'_i}}\rb
 | 0\rangle
 \eeq
with $z_{1}+z_{2}=z, n_1+n_2=n$, $0\leq z_{1}\leq z, 0\leq n_1\leq
n$.

The total wave function can be expanded into the strong coupling
basis,
 \beq
 |IM\rangle = \sum_{K\varphi}c_{K\varphi}|IMK\varphi\rangle,\label{eq:expansion}
 \eeq
with
 \beqn
 &&|IMK\varphi\rangle
   =
   \frac{1}{\sqrt{2(1+\delta_{K0}\delta_{\varphi,\overline{\varphi}})}} \lb |IMK\rangle |\varphi\rangle
  +(-1)^{I-K} |IM -K\rangle|\overline{\varphi}\rangle\rb, \label{eq:basis}
 \eeqn
where $|IMK\rangle$ denotes the Wigner functions
$\sqrt{\frac{2I+1}{8\pi^2}}D^I_{MK}$ and $\varphi$ is a shorthand
notation for the configurations in Eq.~(\ref{eq:configurations}).
The basis states are symmetrized under the point group $D_2$, which
leads to $K-\frac{1}{2}(z_1-z_2)-\frac{1}{2}(n_1-n_2)$ being an even
integer with $\Omega=\cdots, -3/2, 1/2, 5/2, \cdots$. The detailed
explanation for such a restriction is given in Appendix. The reduced
transition probabilities $B(M1)$ and $B(E2)$ can be obtained from
the wave function of PRM with the $M1$ and $E2$
operators~\cite{Zhang07}.

\section{Results and Discussion}

In the PRM calculations for the doublet bands in $^{103,105}$Rh, the
configuration $\pi g^{-1}_{9/2}\otimes\nu
h^{2}_{11/2}$~\cite{Timar04,Timar06} is adopted. By applying the
microscopic self-consistent triaxial relativistic mean field
approach~\cite{MengJ06}, the deformation parameters for ground
states of $^{103,105}$Rh have been obtained with PK1 parameter
set~\cite{MengJ06}, i.e., $\beta=0.235$ and $\gamma=20.6^{\circ}$
for $^{103}$Rh while $\beta=0.228$ and $\gamma=21.1^{\circ}$ for
$^{105}$Rh. Accordingly, $C_p$ and $C_n$ in single-$j$
Hamiltonian~(\ref{eq:hsp}) take values of $-0.43$ and $0.35$ MeV for
$^{103}$Rh while  $-0.42$ and $0.34$ MeV for $^{105}$Rh
~\cite{WangSY08}. For all doublet bands investigated, the moment of
inertia ${\cal J}_0=21.0$ MeV$/\hbar^2$ is adjusted to the
experimental energy spectra. For the electromagnetic transition, the
empirical intrinsic quadrupole moment
$Q_0=(3/\sqrt{5\pi})R_0^2Z\beta\approx2.5$ eb, gyromagnetic ratios
$g_R=Z/A=0.44$, and $g_p(g_{9/2})=1.26$, $g_n(h_{11/2})=-0.21$ are
adopted. The proton and neutron g factors are determined from
$g_{p(n)}=g_l+(g_s-g_l)/(2l+1)$ with $g_l=1(0)$ for proton (neutron)
and $g_s$ =0.6$g_s$(free).

\subsection{$^{103}$Rh}

\begin{figure}[ht]
\includegraphics[width=13cm, bb=0 0 320 240]{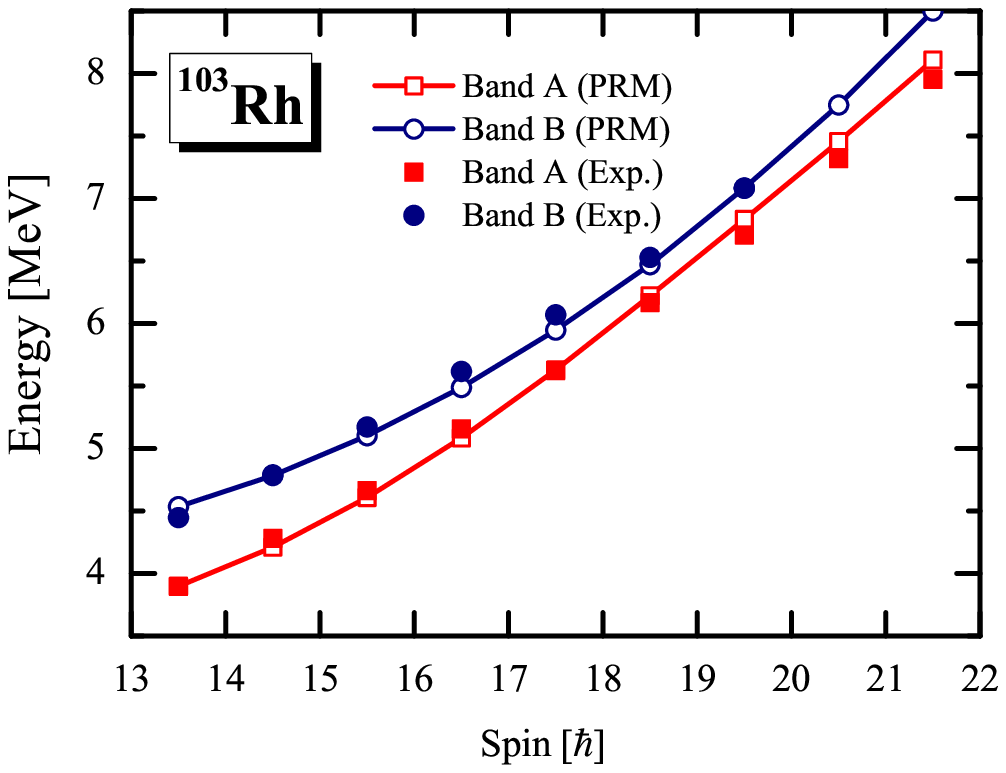}
\caption{\label{fig:energy103} (color online) The energies $E(I)$
for the chiral doublet bands in $^{103}$Rh calculated by the
triaxial PRM with configuration $\pi g^{-1}_{9/2}\otimes \nu
h_{11/2}^2 $ (open symbols connected by full line) in comparison
with the data (filled symbols)~\cite{Timar06}. }
\end{figure}

The calculated excitation energy spectra $E(I)$ for the doublet
bands (denoted by A and B) in $^{103}$Rh are presented in
Fig.~\ref{fig:energy103}, in comparison with the corresponding data
available, i.e. bands 3 and 4 in Ref.~\cite{Timar06}. The
experimental energy spectra are well reproduced by the PRM
calculation with an accuracy within 150 keV. In particular, the
trend and amplitude of the energy separation between the doublet
bands are excellently reproduced. The least energy separation
between the doublet bands happens at $I = 37/2$ for both the
experimental and theoretical energy spectra.

\begin{figure}[t!]
\includegraphics[width=13cm, bb=0 0 320 240]{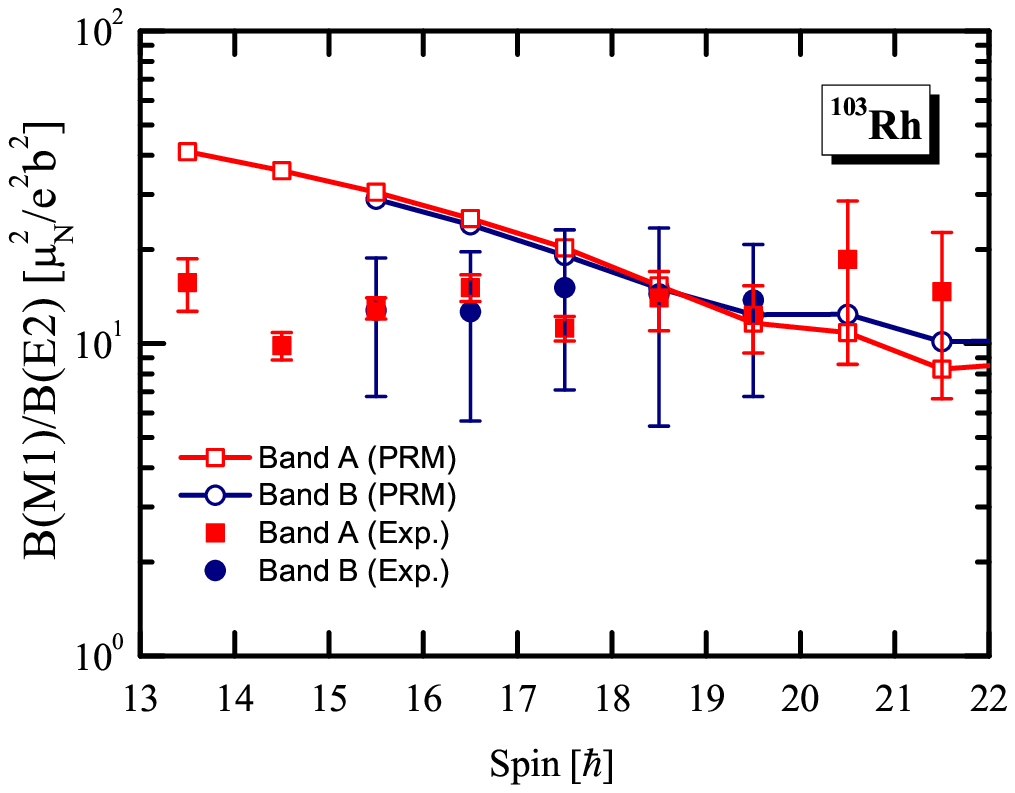}
\caption{\label{fig:M1E2103} (color online) The $B(M1)/B(E2)$ values
calculated by PRM for the chiral doublet bands in $^{103}$Rh (open
symbols connected by full line) in comparison with the data (filled
symbols)~\cite{Timar06}. }
\end{figure}

The calculated in-band $B(M1)/B(E2)$ ratios for the doublet bands in
$^{103}$Rh are presented in Fig.~\ref{fig:M1E2103}, together with
the available data~\cite{Timar06}. For the whole spin region, the
observed in-band $B(M1)/B(E2)$ ratios in the two bands are almost
the same. These features are well reproduced by the PRM calculation.
No strong odd-even staggering of the $B(M1)/B(E2)$ ratios are
observed in the doublet bands, as well as the PRM results. In
Ref.~\cite{Qi0902}, it is pointed out that this odd-even staggering
associates strongly with the characters of nuclear chirality, i.e.,
the staggering is weak in the chiral vibration region while strong
in the static chirality region based on the calculations of  one
particle one hole coupled with a triaxial rotor.

\begin{figure}[t!]
\includegraphics[width=8cm, bb=0 20 340 440]{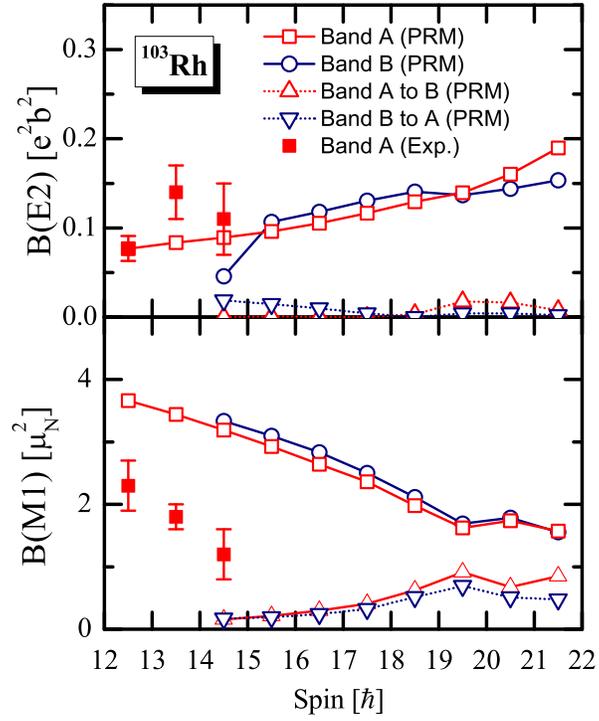}
\caption{\label{fig:M1andE2} (color online) The  $B(M1)$ and $B(E2)$
values calculated by  PRM  for the chiral doublet bands in
$^{103}$Rh (open symbols) in comparison with the data available
(filled symbols)~\cite{Suzuki08}. }
\end{figure}

Lifetime measurements for candidate chiral bands in $^{103,104}$Rh
have been performed in Ref.~\cite{Suzuki08} and the $B(M1)$ and
$B(E2)$ values in band A are extracted. The $B(M1)$ and $B(E2)$
values calculated by means of the PRM for the chiral doublet bands
in $^{103}$Rh are plotted in Fig.~\ref{fig:M1andE2} in comparison
with the data~\cite{Suzuki08}. The predicted $B(E2)$ values increase
smoothly as the spin increasing, while the predicted $B(M1)$ values
decrease smoothly as the spin increasing. The PRM calculations well
reproduce the observed $B(E2)$ values, while somewhat overestimate
the observed $B(M1)$ values. Additionally, both the calculated
in-band $B(M1)$ and $B(E2)$ values are much larger than the
interband ones, which is consistent with the present experimental
observations.

The deviation between the calculated $B(M1)$ values and the observed
data may be understood by the incorporation of one quasiparticle
configuration into the first several states of the observed three
quasiparticles band, as one quasiparticle band gives small $B(M1)$
values. If the pairing interactions are included to treat with the
mixing between one- and three-quasiparticle configurations, the
observed $B(M1)$ values would be better reproduced. In addition, the
dependence of the $B(M1)$ values on $\gamma$ has been checked. It is
found that the $B(M1)$ values of the yrast band for the lower spins
have hardly changed for different $\gamma$ values. The agreement
between the calculated results and $B(M1)$ data does not improve
much by adjusting of $\gamma$.

\begin{figure}[t!]
\includegraphics[width=13cm,  bb= 15 15 333 260]{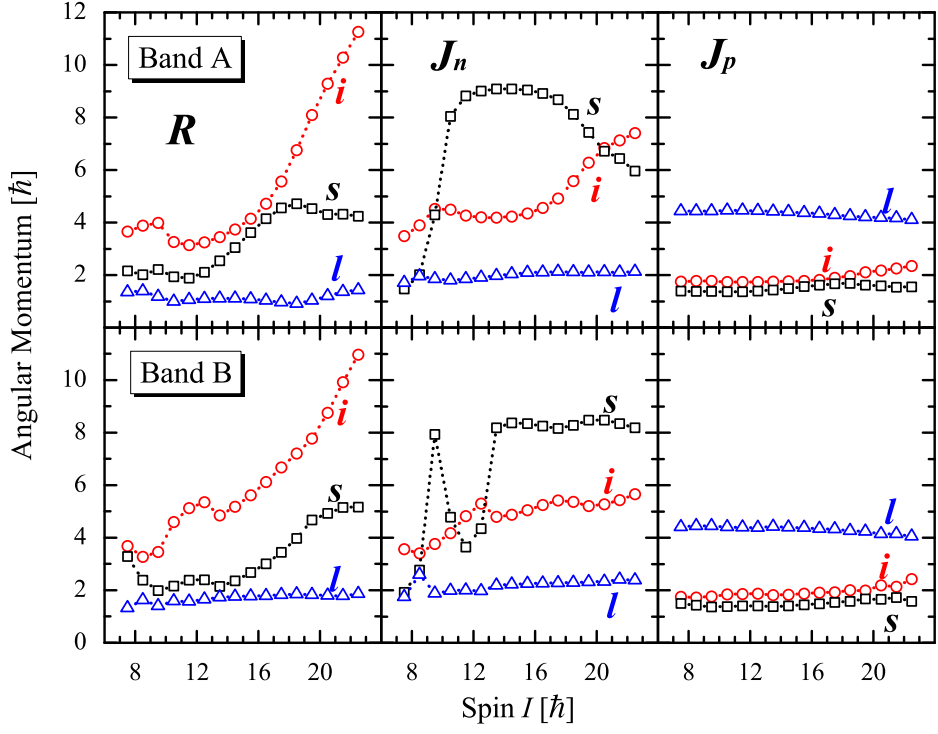}
\caption{\label{fig:spin1} (color online) The root mean square
components along the intermediate ($i$-, circles), short ($s$-,
squares) and long ($l$-, triangles) axis of the core ${R_k} =
\sqrt{\langle \hat{R}_{k}^2 \rangle}$, valence proton $J_{pk} =
\sqrt{\langle \hat{j}_{pk}^2 \rangle}$, and valence neutrons angular
momenta $J_{nk}=\sqrt{\langle (\hat{j}_{(n1)k}+\hat{j}_{(n2)k})^2
\rangle}$ calculated as functions of spin $I$ by means of the PRM
for the doublet bands in $^{103}$Rh. }
\end{figure}

The success in reproducing the energy spectra and transition
probabilities of the doublet bands A and B in $^{103}$Rh suggests
that the PRM calculation correctly accounts for the structure of the
states. To exhibit their chiral geometry hold by the angular
momenta, the rms values of the angular momentum components for the
core ${R_k} = \sqrt{\langle \hat{R}_{k}^2 \rangle}$, the valence
proton $J_{pk} = \sqrt{\langle \hat{j}_{pk}^2\rangle}$, and the
valence neutrons
$J_{nk}=\sqrt{\langle(\hat{j}_{(n1)k}+\hat{j}_{(n2)k})^2\rangle}$ ,
are presented in Fig.~\ref{fig:spin1}, in which $k = i, l, s$,
represent the intermediate, short and long axes respectively.

As shown in Fig.~\ref{fig:spin1}, for both bands A and B with spin
larger than the band head $27/2\hbar$ of band B, the collective core
angular momentum mainly aligns along the intermediate axis. The
angular momentum of the $g_{9/2}$ valence proton hole mainly aligns
along the long axis and that of the two aligned $h_{11/2}$ valence
neutrons mainly along the short axis, which correspond to the
orientation preferred by their interaction with the triaxial core
\cite{FM97}. To be more precise, ${\bf J_{p}} \sim5\hbar$ along
$l$-axis, and both $\bf{J_n}$ $\sim10\hbar$ and $\bf{R}$
$\sim6-12\hbar$ lie in the plane spanned by $i$- and $s$- axis,
which together form the chiral geometry of aplanar rotation.

As the total angular momentum increases, $\bf{R}$ increases
gradually. Due to the Coriolis coupling and the rotational
alignment, both $\bf{J_p}$ and $\bf{J_n}$ moves gradually toward the
$i$-axis, the largest rotational axis. The $i$-component for
$\bf{J_n}$ increase more rapidly than that of $\bf{J_p}$, as the two
aligned $h_{11/2}$ valence neutrons provide larger component on
$i$-axis. At spin $I=37/2$, where the doublet bands have the
smallest energy difference, the orientations of $\bf{R}$, $\bf{J_n}$
and $\bf{J_p}$ for band A and band B are nearly identical. Hence
around this spin the structure comes closest to the ideal chiral
picture of a left- and a right-handed configurations with equal
components $\bf{R}$, $\bf{J_n}$, and $\bf{J_p}$ along the respective
$i$-, $l$-, and $s$-axes.

It is interesting to analyze the evolution of the rms components of
angular momentum before the starting of the doublet bands ($I<
27/2$). For band A, the unpairing process of two $h_{11/2}$ valence
neutrons seems to happen during spin $19/2$ to $23/2$. Before that,
the two valence neutrons are paired and contribute few angular
momenta on three axes. When $I\geq23/2$, the two $h_{11/2}$ neutrons
are unpaired, which mainly align along the $s$-axis (towards
$i$-axis with spin increasing) and contribute $9\hbar$. The onset of
three quasiparticles configuration is pleasingly consistent with the
observed band head ($I=23/2$) of band 3 in Ref.~\cite{Timar04}. For
band B, the unpairing process is complete at spin 27/2, which is
exactly the observed band head of band 4 in Ref.~\cite{Timar04}.
This observation indicates that the present calculation treats
automatically and properly the angular momentum alignment of valence
particles. However, as the moment of inertia is a constant in
present PRM calculation, one cannot expect the simultaneous
quantitative description of 1-qp and 3-qp rotational bands.

\begin{figure}[t!]
 \centering
  \includegraphics[width=13cm, bb=0 115 350 380]{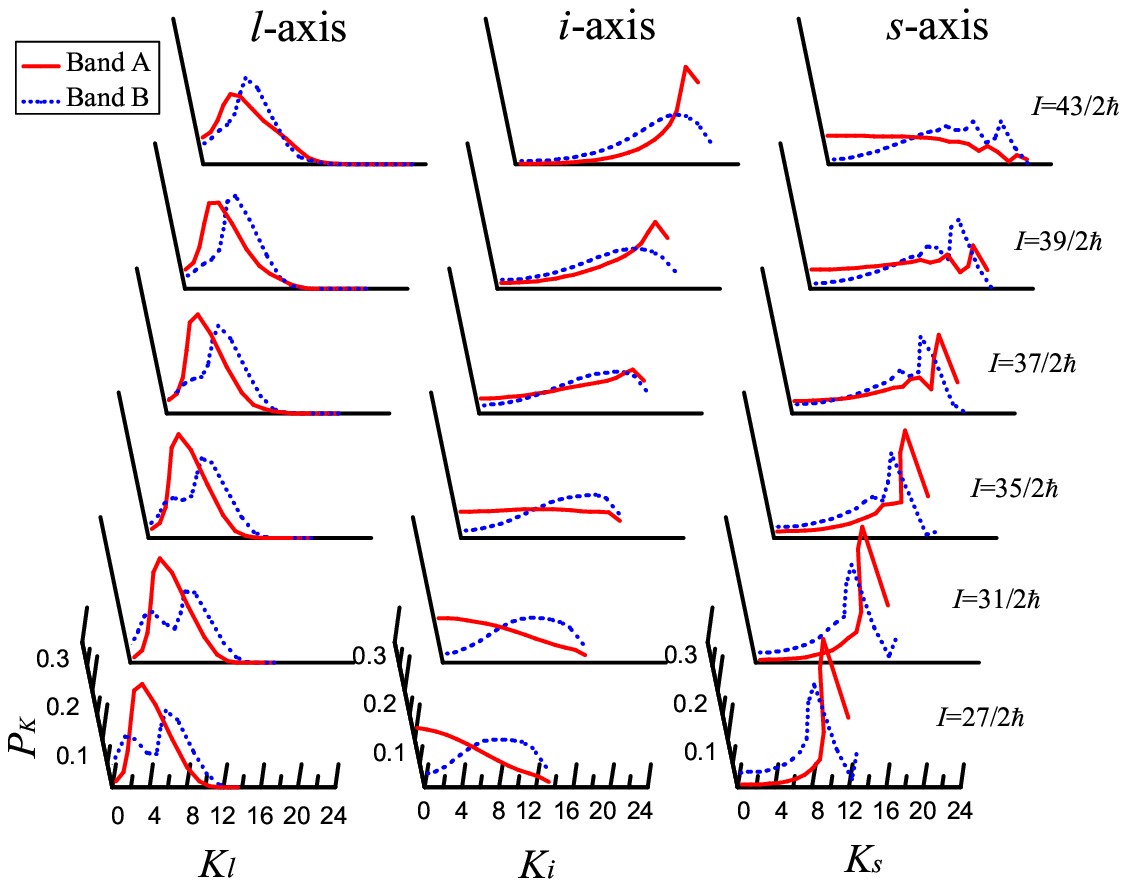}
  \caption{ (color online) The probability distributions for projection of
total angular momentum on the long ($l$-), intermediate ($i$-) and
short ($s$-) axis in PRM for the doublet bands in $^{103}$Rh. }
\label{fig:spin3}
\end{figure}

Further understanding the evolution of the chirality with angular
momentum as performed for the chiral bands in $^{135}$Nd in the
$A\sim130$ region similarly in Ref.~\cite{Qi0901}, the probability
distributions for the projection of the total angular momentum along
the $l$-, $i$- and $s$-axes are given in Fig.~\ref{fig:spin3} for
the doublet bands in $^{103}$Rh. For spin $I=27/2$, near the band
head, the probability distribution of two bands differs as expected
for a chiral vibration. For the lower band A, the maximum
probability for the $i$-axis appears at $K_i=0$, whereas the
probability for the higher band B is zero at $K_i=0$, having its
peak at $K_i=19/2$. These probability distributions resemble those
in the chiral vibration picture shown in TAC+RPA theory, in which
the state (A) and state (B) are considered as the zero phonon and
one phonon states~\cite{Mukhopadhyay07}. The probability
distributions with respect to the $l$-axis have a peak near
$K_l={9/2}$, and respect to the $s$-axis have a peak near $K_s=10$.
Hence the chiral vibration consists in an oscillation of the
collective angular momentum vector $\bf{R}$ through the
$s$-$l$-plane. This reveals the structure of the chiral
vibration~\cite{Mukhopadhyay07}.

At spin $I=37/2$, the probability distributions for band A and B are
very similar. The distributions are peaked at about
$K_l=7,~K_i=17,K_s=16$. The well developed tunneling regime, namely
the well established  static chirality region, is restricted around
$I=37/2$. For higher spin, where the energy difference between the
chiral doublets increases, they attain vibration character again.
This is reflected by the increasing differences between the
probability distributions of bands A and B. In particular the fact
that $P(K_s=0)$ is finite for band A and zero for band B shows that
the motion contains a vibration of the vector ${\bf I}$ through the
$l$-$i$-plane.

\subsection{$^{105}$Rh}

\begin{figure}[th!]
\includegraphics[width=13cm, bb=0 0 320 240]{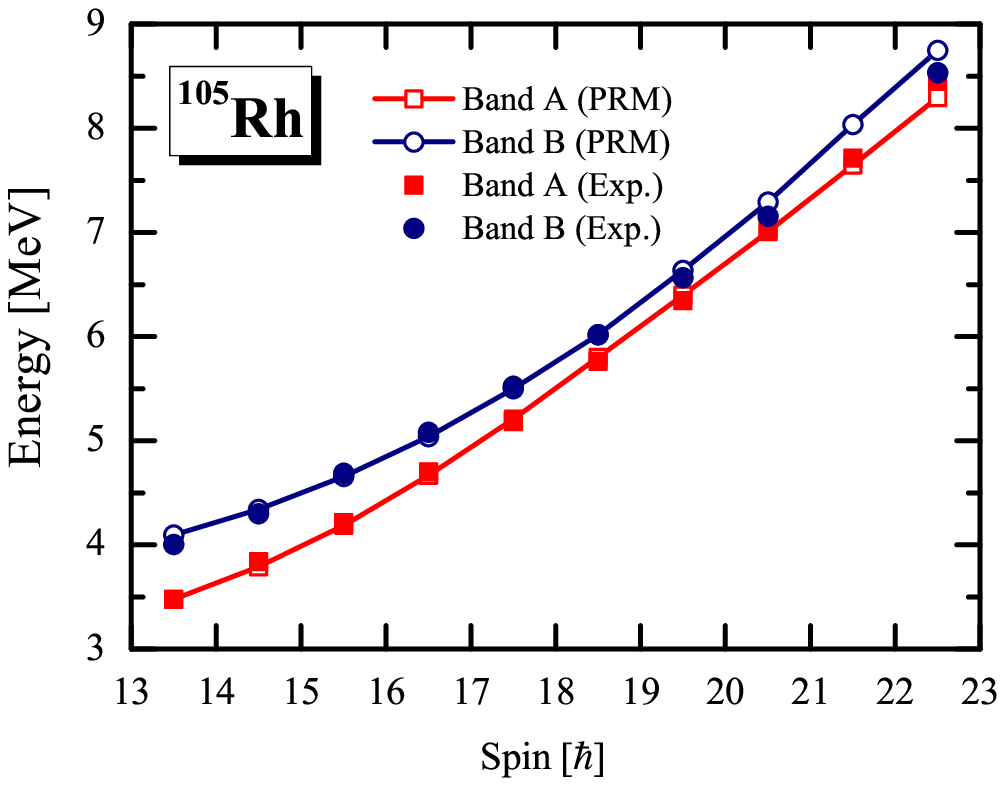}
\caption{\label{fig:energy105} (color online) The energies $E(I)$
for the chiral doublet bands in $^{105}$Rh calculated by the
triaxial PRM with configuration $\pi g^{-1}_{9/2}\otimes \nu
h_{11/2}^2 $ (open symbols connected by full line) in comparison
with the data (filled symbols)~\cite{Alcantara04,Timar04}. }
\end{figure}

The triaxial PRM calculations have also been applied to the doublet
bands with the configuration  $\pi g^{-1}_{9/2}\otimes\nu
h^{2}_{11/2}$ in $^{105}$Rh, namely bands 4 and 5 in
Ref.~\cite{Timar04}. The calculated excitation energy spectra $E(I)$
for the doublet bands A and B in $^{105}$Rh are presented in
Fig.~\ref{fig:energy105}, together with the corresponding
data~\cite{Timar04}. The experimental energy spectra are well
reproduced by the PRM calculation. The calculated results agree with
the data within 100 keV for spin $I<41/2$. The trend and amplitude
for the energy splitting between two doublet bands are also
comparable. However, there still exists certain deviation in energy
splitting that the best degeneracy spin is 41/2 for data, but 39/2
for PRM calculation.

\begin{figure}[h!]
\includegraphics[width=13cm, bb=0 0 320 240]{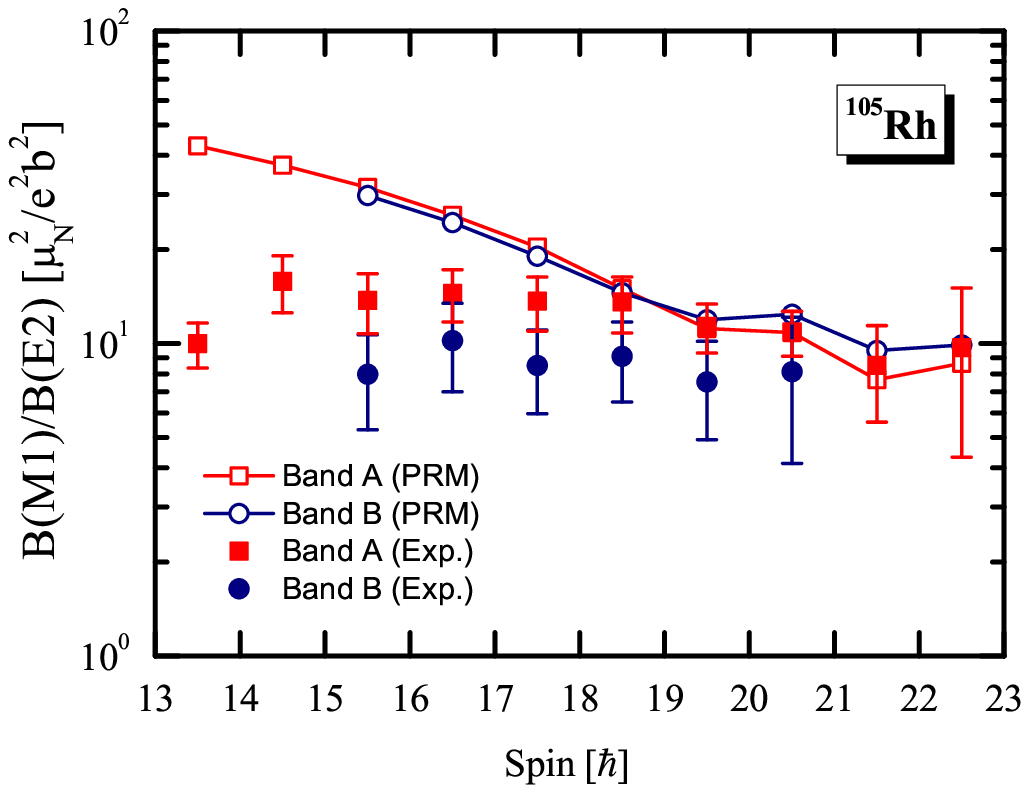}
\caption{\label{fig:M1E2105} (color online) The  $B(M1)/B(E2)$
values calculated by  PRM  for the chiral doublet bands in
$^{105}$Rh (open symbols connected by full line) in comparison with
the data (filled symbols)~\cite{Alcantara04,Timar04}. }
\end{figure}

\begin{figure}[h!]
\includegraphics[width=8cm, bb=0 20 340 440]{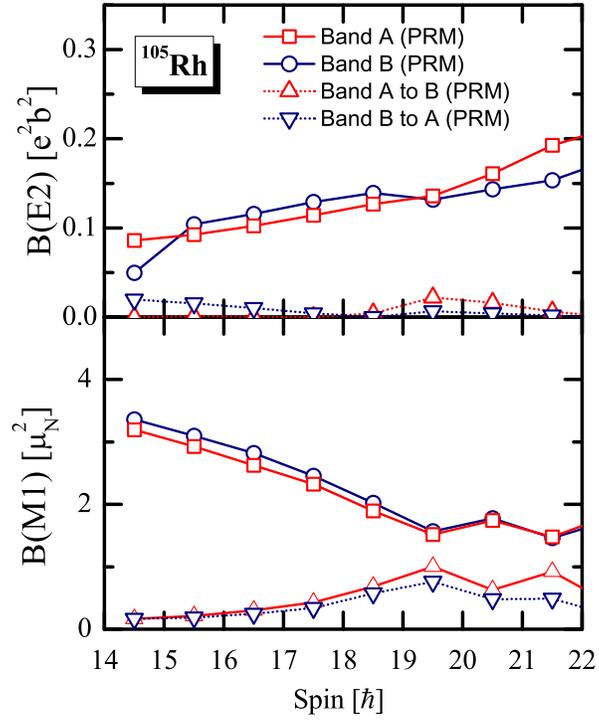}
\caption{\label{fig:M1andE2105} (color online) The  $B(M1)$ and
$B(E2)$ values calculated by PRM  for the chiral doublet bands in
$^{105}$Rh. }
\end{figure}

The calculated in-band $B(M1)/B(E2)$ ratios for the doublet bands in
$^{105}$Rh are presented in Fig.~\ref{fig:M1E2105}, together with
the available data~\cite{Timar06}. For the whole spin region, the
observed in-band $B(M1)/B(E2)$ ratios in the two bands are similar
within error bar. These features are reproduced  by the PRM
calculation. The small difference of magnitude can be seen in
experimental data as well at large spins for PRM calculations. The
staggering of the $B(M1)/B(E2)$ ratios are not obvious in this
nucleus, which is consistent with the PRM calculation. The absolute
$B(E2)$ and $B(M1)$ values for the doublet bands in $^{105}$Rh from
the PRM calculation are also plotted in Fig.~\ref{fig:M1andE2105},
in order to compare with future lifetime data.

To study their chiral geometry, the rms values of the angular
momentum components for the core, the valence proton and the valence
neutrons,  and the probability distributions of projection $K$ of
total angular momentum are also calculated, and the results are
almost the same as that in $^{103}$Rh. Thus the evolution of the
chiral geometry with angular momentum in $^{105}$Rh are almost
identical to that in $^{103}$Rh. To avoid repetition, these results
are not shown here.

It is interesting to note that in $^{105}$Rh, a pair of negative
parity bands (band 7 and 8 in Ref.~\cite{Alcantara04}) have been
also regarded as another chiral doublet bands~\cite{Alcantara04}.
The corresponding quasiparticle configuration was suggested as $\pi
g^{-1}_{9/2}\otimes\nu h_{11/2}(g_{7/2}, d_{5/2})$. This fact
indicates the possible experimental evidence for the existence of
M$\chi$D~\cite{MengJ06} in $^{105}$Rh, namely a pair of chiral bands
with positive parity and configuration $\pi g^{-1}_{9/2}\otimes \nu
h_{11/2}^2$, while another pair of chiral bands with negative parity
and configuration $\pi g^{-1}_{9/2}\otimes\nu h_{11/2}(g_{7/2},
d_{5/2})$. As the latter configuration involves the orbits of
pseudospin doublet states $(g_{7/2}, d_{5/2})$, a competing
interpretation of band 7 and 8 in Ref.~\cite{Alcantara04} includes
the pseudospin doublet bands~\cite{Meng10}. Further efforts are
needed to address this point.

\subsection{The influence of $\gamma$ on the doublet bands}

\begin{figure}[t!]
\includegraphics[width=13cm, bb= 20 20 450 360]{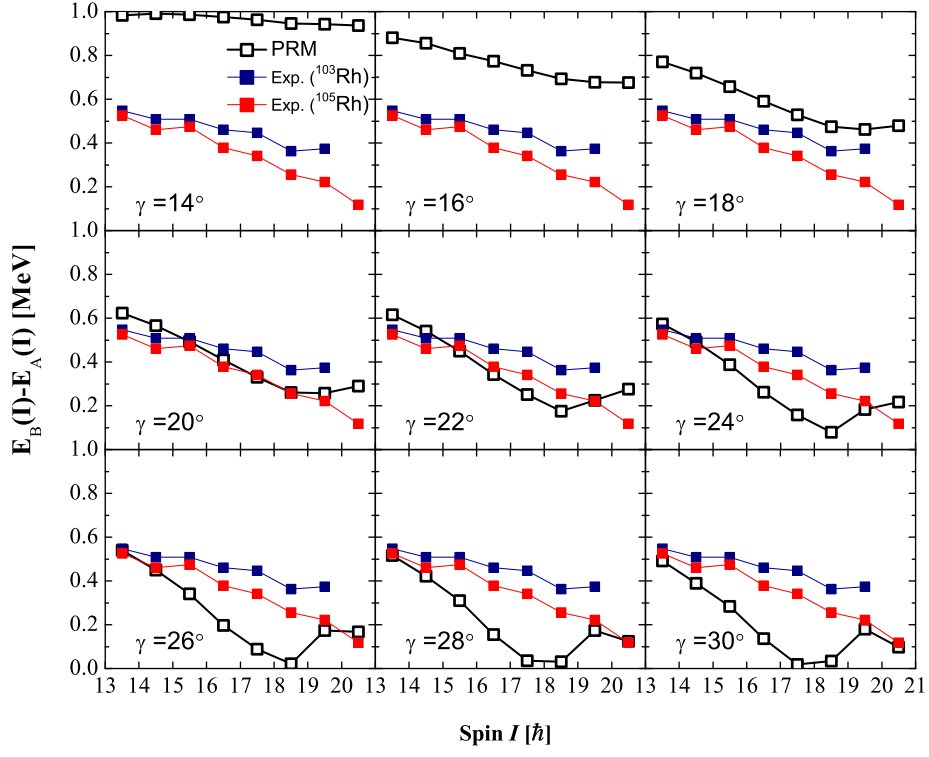}
\caption{\label{fig:DEI} (color online) Energy difference between
the two lowest bands A and B calculated in PRM with different
triaxiality parameters $\gamma$ with configuration $\pi
g^{-1}_{9/2}\otimes \nu h_{11/2}^2 $ (open symbols) in comparison
with the data of $^{103}$Rh and $^{105}$Rh (filled symbols). }
\end{figure}

\begin{figure}[th!]
\includegraphics[width=13cm, bb= 20 20 450 360]{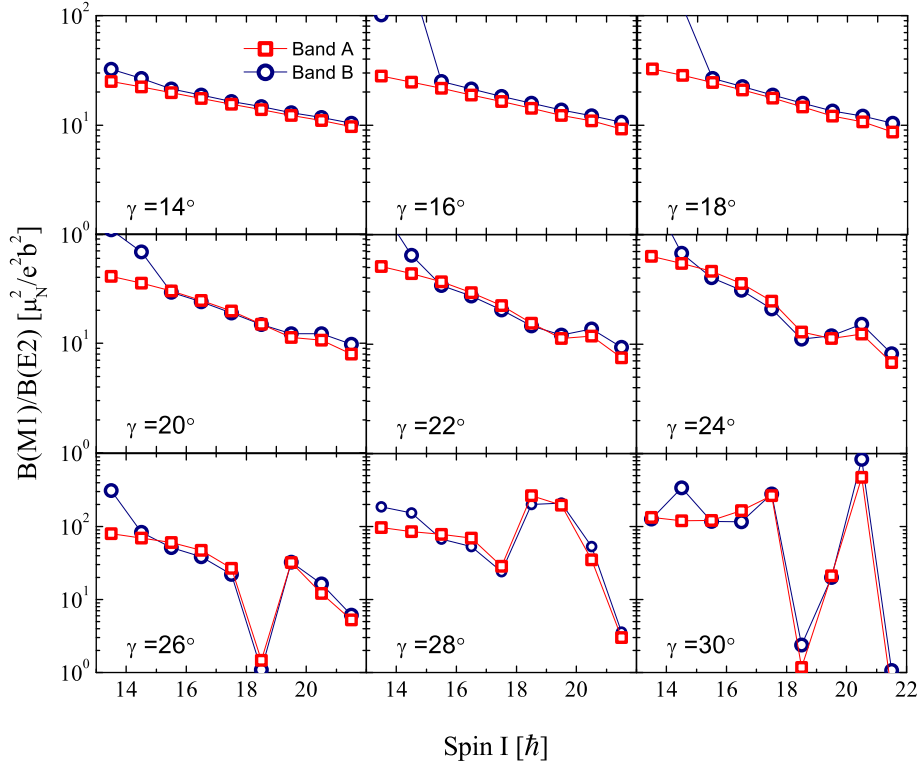}
\caption{\label{fig:EM} (color online) The in-band $B(M1)/B(E2)$
values of the two lowest bands A and B calculated in PRM with
different triaxiality parameters $\gamma$ with configuration $\pi
g^{-1}_{9/2}\otimes \nu h_{11/2}^2 $ (open symbols).}
\end{figure}

In the previous study of chiral doublet bands in odd-odd nuclei with
particle rotor model, it has been found that the triaxial
deformation $\gamma$ is one of the most sensitive parameters for the
properties of the doublet bands~\cite{Qi0902}. Here, we made a
systematic triaxial PRM calculation with configuration $\pi
g^{-1}_{9/2}\otimes \nu h_{11/2}^2 $, and plot the energy difference
between the two lowest bands A and B and their in-band $B(M1)/B(E2)$
ratios for different triaxiality parameter $\gamma$ in
Fig.~\ref{fig:DEI} and~\ref{fig:EM}.

As shown in Fig.~\ref{fig:DEI}, as the $\gamma$ value increases, the
calculated energy difference between doublet bands decreases
rapidly. The smallest energy differences are obtained at spin 37/2
for most $\gamma$ values.  At this point, the energy differences are
946, 693, 475, 261, 175, 79, 21, 30, 34 keV for $\gamma=14^{\circ},
16^{\circ}, 18^{\circ}, 20^{\circ}, 22^{\circ}, 24^{\circ},
26^{\circ}, 28^{\circ}, 30^{\circ},$ respectively. It is found that
around $\gamma=20^{\circ}$, the calculated energy difference agrees
with the data in $^{103,105}$Rh very well.

As shown in Fig.~\ref{fig:EM}, the calculated in-band $B(M1)/B(E2)$
ratios in the doublet bands are almost the same over the whole spin
region for different triaxiality parameters $\gamma$. These
calculated in-band $B(M1)/B(E2)$ values are sensitive to the
triaxiality parameter $\gamma$. Such two pronounced features for
$B(M1)/B(E2)$, i.e., the similarity between the doublet bands  and
the sensitivity to $\gamma$, are consistent with the features in
chiral doublet bands calculated with 1-particle-1-hole coupled to
triaxial rotor model in Ref.~\cite{Qi0902}. However, the staggering
of the $B(M1)/B(E2)$ ratios are not obvious for $\gamma<24^\circ$,
and the staggering appear only around spin $37/2$ for
$\gamma=26^\circ, 28^\circ, 30^\circ$, which are very different from
the features in the case of 1-particle-1-hole configurations in
Ref.~\cite{Qi0902}.

\section{Summary}
In summary, adopting the particle-rotor model, which couples a
triaxial rotor with one valence proton hole and two valence
neutrons, the candidate chiral doublet bands with $\pi
g_{9/2}^{-1}\otimes\nu h^{2}_{11/2}$ configuration in the odd-$A$ Rh
isotopes, $^{103}$Rh and  $^{105}$Rh, are investigated. The
agreement is excellent for the energy spectra of the doublet bands
over entire spin range and in-band $B(M1)/B(E2)$ ratios at higher
spin range. The absolute $B(M1)$ and $B(E2)$ transition
probabilities for the doublet bands in $^{103}$Rh and $^{105}$Rh
have been presented, which has reasonable agreement with the data
available and invites further lifetime measurements in the chiral
rotation region.

Both in $^{103}$Rh and  $^{105}$Rh, the chiral doublet bands  start
as a chiral vibration of the angular momentum about the intermediate
axis. Static chirality is reached at $I=37/2$, where the two bands
approach each other closest. After that the two bands again develop
into a chiral vibration of the angular momentum about the short
axis.

For the doublet bands with $\pi g_{9/2}^{-1}\otimes\nu h^{2}_{11/2}$
configuration, it is found that both the energy spectra difference
and their in-band $B(M1)/B(E2)$ ratios are sensitive to the
triaxiality parameter $\gamma$. No regular staggering of the
$B(M1)/B(E2)$ ratios is found for the doublet bands with this
configuration for different $\gamma$ values.

\section*{Acknowledgements}
This work is partly supported by the National Natural Science
Foundation of China under Grant Nos. 11005069, 10975007, 10975008,
10875074, 10775004, the Independent Innovation Foundation of
Shandong University (IIFSDU), and the Major State Research
Development Program of China (No. 2007CB815000).

\section{Appendix}

In the appendix, we will show the restriction of the model basis
Eq.~(\ref{eq:basis}). The model basis should be symmetrized under
the $D_2$ point group, namely the operators $\hat{R}_1(\pi)$,
$\hat{R}_2(\pi)$ and $\hat{R}_3(\pi)$ which represent rotation of
$180^\circ$ about the three principal axes in body-fixed frame. When
the model basis is written as Eq.~(\ref{eq:basis}), it has been
ensured that the basis is invariant under $\hat{R}_2(\pi)$ (see such
as Refs.~\cite{Carlsson06, Ragnarsson88}).

For
$\hat{R}_3(\pi)=exp[\frac{-i}{\hbar}\pi(\hat{I}_3-\hat{j}_{3})]$,
 \beqn \hat{R}_3(\pi)a^\dag_{\nu}|0\rangle &=& \hat{R}_3(\pi) \sum_{\alpha
 \Omega}c_{\alpha \Omega}^{(\nu)}  |\alpha,\Omega \rangle=
 \sum_{\alpha \Omega}c_{\alpha \Omega}^{(\nu)} (-1) ^{-\Omega}
 |\alpha,\Omega \rangle.
 \eeqn
Because $\Omega$ has been restricted as $\cdots, -3/2, 1/2, 5/2,
\cdots$ in Eq.~(\ref{eq:spwf}),
 \beqn
 R_3(\pi)a^\dag_{\nu}|0\rangle  &=& \sum_{\alpha \Omega}c_{\alpha
\Omega}^{(\nu)} (-1) ^{-(1/2+2*integer)} |\alpha,\Omega \rangle
=(-1)^{-1/2} a^\dag_{\nu}|0\rangle. \eeqn
 Similarly, $
R_3(\pi){a}^{+}_{\overline{\nu}}| 0 \rangle =(-1)^{1/2}
{a}^{+}_{\overline{\nu}}| 0 \rangle. $ Hence, when $\hat{R}_3(\pi)$
operates on the intrinsic wave function, we get
 \beqn\label{eq:ap02}
 R_3(\pi) |\varphi\rangle &=&R_3(\pi) \lb\prod_{i=1}^{z_{1}}a^\dag_{p, \nu_i}\rb
 \lb\prod_{i=1}^{z_{2}}a^\dag_{p,\overline{\mu_i}}\rb
 \lb\prod_{i=1}^{n_{1}}a^\dag_{n,\nu'_i}\rb
 \lb\prod_{i=1}^{n_2}a^\dag_{n,\overline{\mu'_i}}\rb
 | 0\rangle\nonumber\\
 &=&(-1)^{-\frac{1}{2}z_1}(-1)^{\frac{1}{2}z_2}(-1)^{-\frac{1}{2}n_1}(-1)^{\frac{1}{2}n_2}|\varphi\rangle.
 \eeqn
Consequently,
 \beqn
 R_3(\pi)|IMK\rangle |\varphi\rangle &=&(-1)^{K-\frac{1}{2}(z_1-z_2)-\frac{1}{2}(n_1-n_2)}|IMK\rangle
 |\varphi\rangle;
\\
 R_3(\pi)|IM-K\rangle |\overline{\varphi}\rangle &=&(-1)^{-K+\frac{1}{2}(z_1-z_2)+\frac{1}{2}(n_1-n_2)}|IM-K\rangle
 |\overline{\varphi}\rangle.
 \eeqn
 To ensure the model basis $|IMK\varphi\rangle$ in Eq.~(\ref{eq:basis}) is symmetrized under
$\hat{R}_3(\pi)$,  $K-\frac{1}{2}(z_1-z_2)-\frac{1}{2}(n_1-n_2)$
must be an even integer.

According to the group theory, if the model basis
$|IMK\varphi\rangle$ is symmetrized under $\hat{R}_2(\pi)$ and
$\hat{R}_3(\pi)$, it can be ensured that the  basis is also
symmetrized under $\hat{R}_1(\pi)$, namely symmetrized under the
$D_2$ point group.


\end{CJK*}
\end{document}